# Ultra-Broadband Acoustic Metasurface for Manipulating the Reflected Waves

Yi-Fan Zhu[1,2,3†], Xin-Ye Zou[1,2,3†], Rui-Qi Li[1], Xue Jiang[1], Juan Tu[1], Bin Liang[1,2,3*], and Jian-Chun Cheng[1,2,3*]

[1]*Key Laboratory of Modern Acoustics, MOE, Institute of Acoustics, Department of Physics, Nanjing University, Nanjing* 210093, *P. R. China*

[2]*Collaborative Innovation Center for Advanced Microstructures, Nanjing University,* 210093, *P. R. China*

[3]*State Key Laboratory of Acoustics, Chinese Academy of Sciences, Beijing* 100190, *P. R. China*

Abstract

We have designed and experimentally realized an ultra-broadband acoustic metasurface (UBAM) capable of going beyond the intrinsic limitation of bandwidth in existing designs of optical/acoustical metasurfaces. Both the numerical and experimental results demonstrate that the UBAM made of subwavelength gratings can manipulate the reflected phase-front within a bandwidth larger than 2 octaves. A simple physical model based on the phased array theory is developed for interpreting this extraordinary phenomenon as well as obtaining deeper insight to the underlying physics of our design. We anticipate the UBAM to open new avenue to the design and application of broadband acoustical devices.

PACS numbers: 81.05.Xj, 43.20.+g, 46.40.Cd

Recently, the emerging field of metasurfaces, artificially designed surfaces with sub-wavelength structures in optics, has generalized the Snell's law [1] and led to various unconventional phenomena [2-12] with the novel idea of introducing spatial variations in the phase response at the interface. The realization of the acoustic counterpart of metasurface is highly desired for all the acoustic-wave-based applications ranging from noninvasive tests to medical ultrasound imaging/diagnosis. Considerable interests also exist for broadband metasurfaces with particular importance in practical situations generally involving multi-frequency signals. An effective structure of broadband acoustic metasurface providing opportunity to achieve both purposes, however, still remains challenging despite the substantial significance for both physics and engineering communities. So far, the acoustic metasurfaces are less investigated experimentally [13-15]. Moreover, the previous designs of metasurfaces, both in optics and acoustics, inherently have narrow bandwidths, primarily due to two aspects of reasons: First, to realize phase jumps, periodic arrays of resonance elements are usually used that have to be frequency-sensitive, e.g., optical antenna [1,3] or acoustical coiling or Helmholtz structures [13-15]. Second, the transverse dimensions of optical/acoustical arrays at interface must depend on the wavelength to obtain the desired phase gradient. An intrinsically distinct mechanism is to be presented to overcome the frequency limitation in the metasurface design and achieve acoustic manipulation to an unprecedented level.

In this Letter, we advance the idea of acoustic metasurface by proposing the concept of ultra-broadband acoustic metasurface (UBAM). As a representative example, a distinct property of yielding the extraordinary reflection by two-dimensional (2-D) UBAM structures has been designed theoretically and realized experimentally. Excellent agreement is observed between theoretical predictions and experimental results, both showing that the metasurface made of subwavelength gratings with a non-periodic pattern can manipulate the reflected wave without frequency limitation. A simple model based on the phased array theory is developed to interpret this extraordinary phenomenon, which provides a deeper insight to the underlying physics mechanism of our design. We have also discussed the potential of the proposed scheme to realize various unconventional manipulations such as abnormal focusing and non-diffractive beams generation. With the hitherto inaccessible capability of breaking the conventional physical perception and intuition with respect to the planar appearance, the realization of UBAM may push more invention possibilities of broadband acoustic components and broaden the applications of metasurface-based devices.

The design of an UBAM model starts from the generalized law of reflection [1]

$$\sin(\theta_r) - \sin(\theta_i) = \frac{\lambda}{2\pi}\frac{d\phi}{dx} \quad , \qquad (1)$$

where $\theta_r$ and $\theta_i$ are the angles of reflection and incidence, respectively, and $\phi$ is the phase factor. The reflected angle $\theta_r$ can be calculated as

$$\theta_r = \arcsin\left[\sin(\theta_i) + \frac{\lambda}{2\pi}\frac{d\phi}{dx}\right] \quad . \qquad (2)$$

Equation (2) implies that the incident wave front can be reflected along arbitrary direction when a suitable gradient of phase discontinuity is distributed along the interface. The phase gradient adopted in previous optic and acoustic metasurfaces is designed for a specific frequency, which in turn makes the item $(\lambda/2\pi)(d\phi/dx)$ dependent of incident wavelength. If we adopt a new phase response mechanism to eliminate the wavelength-dependent item, a broadband metasurface can be obtained.

Figure 1(a) schematically depicts 2-D extraordinary reflection by the UBAM (ERUBAM) with certain acoustic grating pattern to yield the desired phase response. Different from existing metasurface with periodic, the ERUBAM structure consists of sub-wavelength grooves with varied depths, denoted as $h_j$ for the *j*th groove. The incident wave will propagate along a round trip in each waveguide-like groove with sub-wavelength width that only allows plane waves to propagate. Then the phase is sequentially delayed by different degrees when passing through these grooves. By directly adjusting the depths of individual grooves, the reflected phase at the grating surface is determined by the total propagation distance, i.e., $\phi_j = -4\pi h_j/\lambda$. For sufficiently small *d*, the phase gradient can be expressed as

$$\frac{d\phi}{dx} = \frac{\phi_j - \phi_{j+1}}{d} = \frac{4\pi(h_{j+1} - h_j)}{\lambda d}, \tag{3}$$

which gives

$$\frac{\lambda}{2\pi}\frac{d\phi}{dx} = \frac{2(h_{j+1} - h_j)}{d}. \tag{4}$$

Equation (2) gives the extraordinary reflected angle

$$\theta_r = \arcsin[\sin(\theta_i) + 2g], \tag{5}$$

where $g = (h_{j+1} - h_j)/d$ is the gradient of the grooves array. It is noted in Eq. (4) that the item $(\lambda/2\pi)(d\phi/dx)$ is frequency-independent and only dominated by the choice of *g*. This is the key to break the frequency limitation and make an UBAM possible. In fact, all we need is a specific phase gradient. At different frequencies, therefore, the phase profiles can be the same despite the variation in the phase response at a specific *x* position.

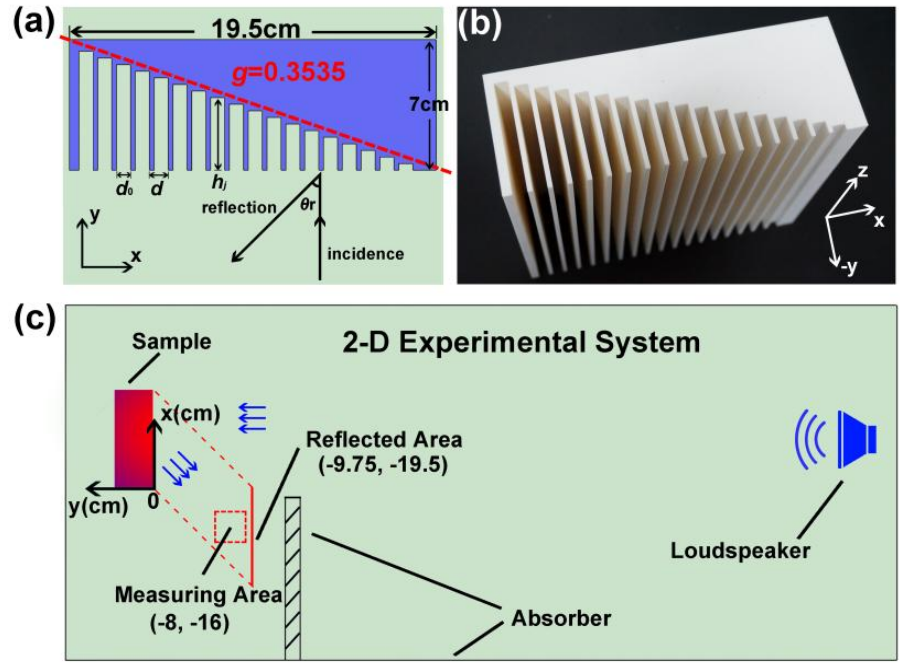

**FIG. 1** (a) Schematic cross-section of ERUBAM grating; (b) Photograph of ERUBAM sample made of acrylonitrile butadiene styrene plastic (ABS) manufactured by 3D printer. The size in *z* direction is 18cm; (c) Schematic of 2-D experimental system. The measuring area is a 6cm×6cm square region. The reflect area is defined as a 19.5cm wide cross profile. The center coordinates of different areas are marked in the figure.

The following analyses take a particular ERUBAM structure with constant gradient of groove arrays as example. As shown in Fig. 1(a), the parameters are: number of grooves $n=18$, period $d=1$cm and width $d_0=0.75$cm. The depth $h_j$ of each groove increases from 3.535mm to 63.63mm with step of 3.535mm ($g = 0.3535$). For an incident wave with a fundamental frequency of $f_0 = c_0/2h_{18} = 2828.4 \text{Hz}$ ($c_0$ is sound velocity in air), the phase range modulated by the eighteen-groove ERUBAM is $0-2\pi$ with step of $\pi/9$. Thus, for an incident wave with arbitrary frequency of $\beta f_0$ ($\beta$ is the normalized frequency scale), the reflected phase can be modulated between $0-2\beta\pi$ with step of $\beta\pi/9$.

Figure 1(b) and 1(c) show the photograph of an ERUBAM sample and the 2-D schematic of the experimental system, respectively. The measurements were performed in the anechoic chamber. To generate an acoustic plane wave, a loudspeaker was located 3m away from the sample. The measuring area and the center of loudspeaker are located in the same *x-y* plane. The reflected angle deduced from Eq. (5) for normally-incident wave is $45^\circ$. An absorptive plate was placed in between to separate the incident and reflected acoustic fields. The error caused by diffraction effect near the edge of the plate should be negligible since both the size of ERUBAM and the measuring area are much larger than the incident wavelength.

To demonstrate the effectiveness of ERUBAM, we have performed a series of measurements on the acoustic pressure distributions of the reflected wave within a broadband, and the typical results will be presented in what follows. Numerical simulations have also been carried out for comparison. Throughout the paper, the numerical simulations are performed by using The Finite Element Method (FEM) based on commercial software COMSOL Multiphysics$^{\text{TM}}$ 4.3. The numerical and the experimental results are illustrated in Fig. 2 for three particular frequencies: $2.5f_0$,

$4.3f_0$, and $6f_0$ (viz., 7.071 kHz, 12.162 kHz and 16.970 kHz). As shown in Fig. 2(a), for $f = 7.071$ kHz, the extraordinary reflection angle can be identified as $45°$, which is in good agreement with the theoretical prediction. Figure 2(b) shows the discontinuous phase distribution at different locations of the interface for $f = 7.071$ kHz. Here, the discrete phase number (PN) in $2\pi$ range is defined as PN=$n/\beta$. The solid line shown in Fig. 2(b) indicates the discrete phase number for $f = 7.071$ kHz is PN=18/2.5=7.2. Similarly, Figs. 2(c) to 2(f) illustrate the acoustic pressure distributions and corresponding discontinuous phase distributions for the cases of $f = 4.3f_0$ and $6f_0$ (viz., 12.162 kHz and 16.970 kHz). The PN values calculated for these two cases are 4.19 and 3, respectively. The results indicate that the extraordinary reflection of $45°$ can be observed within an ultra-broadband frequency range (larger than 2 octaves).

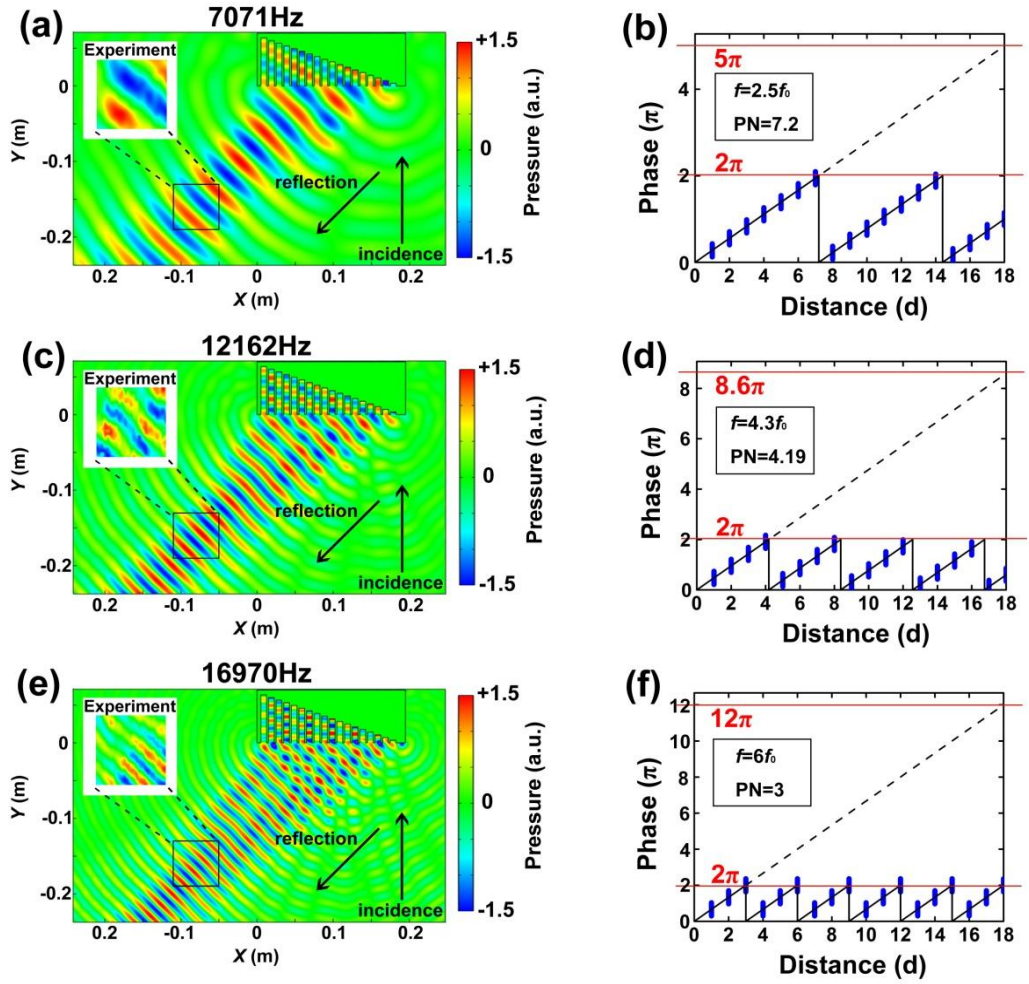

**FIG. 2** Scattering acoustic pressure field of ERUBAM in simulation. The insets show the experimental results of the scattering acoustic pressure field in measuring area: (a) 7.071kHz, (c) 12.162kHz; (e) 16.970kHz. The discontinuous phase distribution along interface with different PNs: (b) PN=7.2 (7.071kHz); (d) PN=4.19 (12.162kHz); (f) PN=3 (16.970kHz).

To reveal the intrinsic physics of UBAM structure, we develop a simple physical model based on the phased array theory to account for the extraordinary characteristics of broadband manipulation. When the incident wave enters each groove, it will induce a vibration of air at orifice of groove. And such a vibration behaves as a secondary radiation source to form abnormal reflected wave. The strength factor $A_j$ of secondary source is estimated approximately by $A_j = Sv_j$, where $v_j$ is normal velocity at the orifice and $S$ is cross-section of each groove. Because of different depth of each groove, the phase of each secondary source is delayed and modulated by the grating. On the other hand, each secondary source can be considered as an acoustic line source radiated to half space due to sub-wavelength width of the groove. Therefore, we can simply consider the abnormal reflected acoustic wave as a new acoustic radiation by a series of secondary sources with different phase delays, as shown in Fig. 3(a). Assuming the distance between observation point and the *j*th point source is $\rho_j$, the total reflected pressure can be expressed as

$$p(\rho,\theta,\omega) = \sum_{j=1}^{N} A_j \exp(i\phi_j) H_0^{(1)}(k_0 \rho_j), \qquad (6)$$

where $\phi_j$ is the phase delay of each secondary sources. When the incident wave is a plane wave and width of each groove is same, all of the strength factors can be considered as equal, i.e., $A_j \equiv A$. In our model, the phase delay by each groove is $\phi_j = 2k_0 h_j$ with $h_j = (N-j)dg$ ($N = n+1 = 19$). Then, at far-field, Eq. (6) can be written as

$$p(\rho,\theta,\omega) \approx A e^{-i\pi/4} \sum_{j=1}^{N} \sqrt{\frac{2}{\pi \rho_j}} \exp\left[i(k_0 \rho_j + 2h_j)\right], \qquad (7)$$

where $\rho_j = \rho_1 - (j-1)\Delta\rho$ and $\Delta\rho = d\sin\theta$, $\rho_1$ is the distance between observation point and the first source. The center of the source array is approximately set as

$\rho = \rho_1 - L\Delta\rho/2d$ as reference point, where $L = (N-1)d$ is the total length of line source array. Then one has

$$p(\rho,\theta,\omega) = A\sqrt{\frac{2}{\pi\rho}} e^{i(k_0\rho - \pi/4)} \frac{\sin[Nk_0 d(\sin\theta - 2g)/2]}{\sin[k_0 d(\sin\theta - 2g)/2]}. \tag{8}$$

It is noteworthy that as predicted by Eq. (8), the maximum of the radiation always appears at a particular angle, $\theta_0 = \arcsin(2g)$, which is totally independent of $k_0$. The field of the extraordinary reflected wave at $\theta_0$ is $p(\rho,\theta_0,\omega) \approx AN e^{i(k_0\rho - \pi/4)}/\sqrt{2/\pi\rho}$. Then Eq. (8) can be expressed as $p(r,\theta,\omega) = p(r,\theta_0,\omega)D(\theta)$, where $D(\theta)$ is the directional factor, given as below

$$D(\theta) = \frac{1}{N} \frac{\sin[Nk_0 d(\sin\theta - 2g)/2]}{\sin[k_0 d(\sin\theta - 2g)/2]}. \tag{9}$$

Figures 3(b)-3(g) show the acoustic pressure fields calculated by Eq. (6) and the corresponding $D(\theta)$ calculated by Eq. (9) at different frequencies. The results suggest that the extraordinary reflection appear at $\theta = 45°$ with no sidelobes, which agree excellently with the numerical and experimental results.

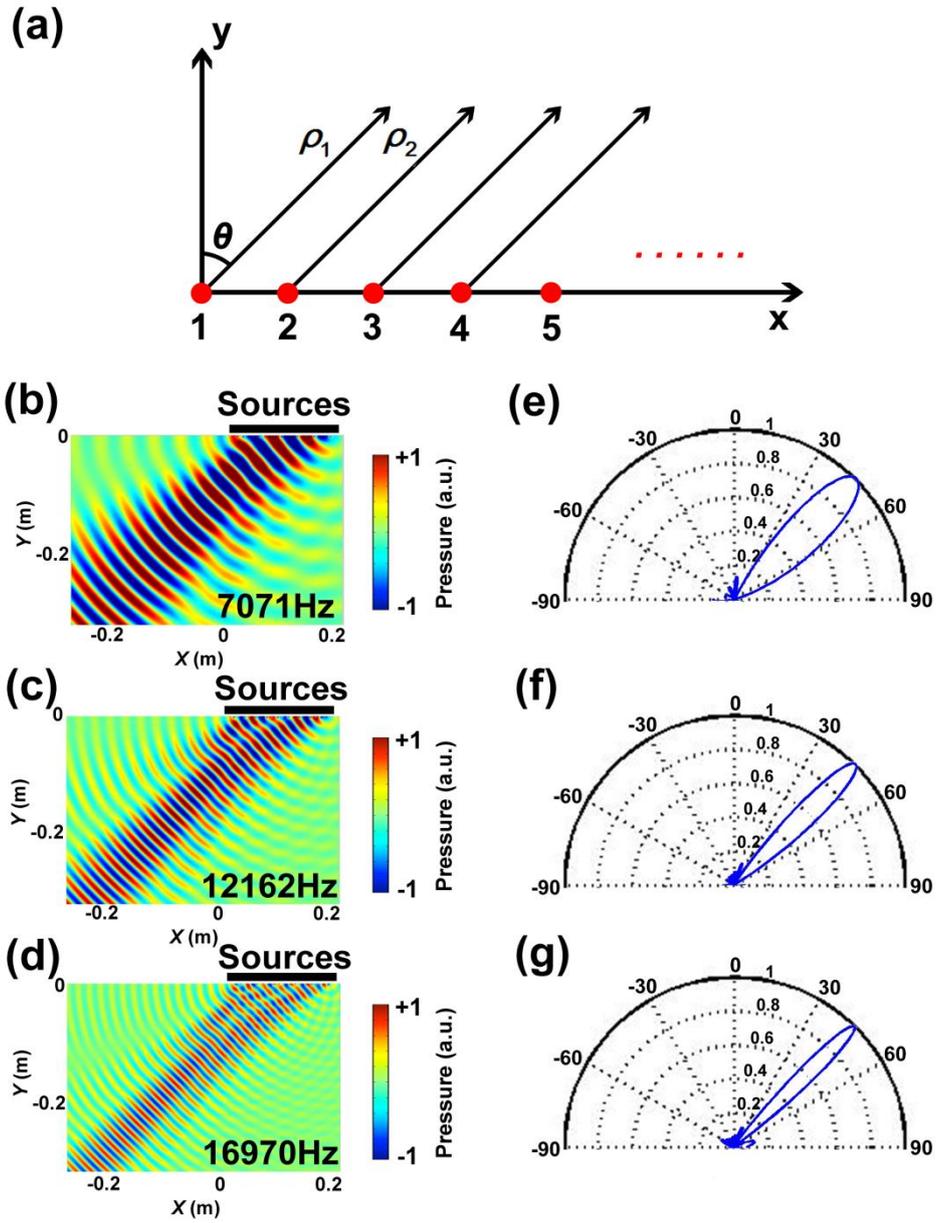

**FIG. 3** (a) A schematic of physical model. The acoustic pressure fields and $D(\theta)$ calculated by Eqs. (6) and (9), respectively, at (b,e) 7.071kHz, (c,f) 12.162kHz, and (d,g) 16.970kHz.

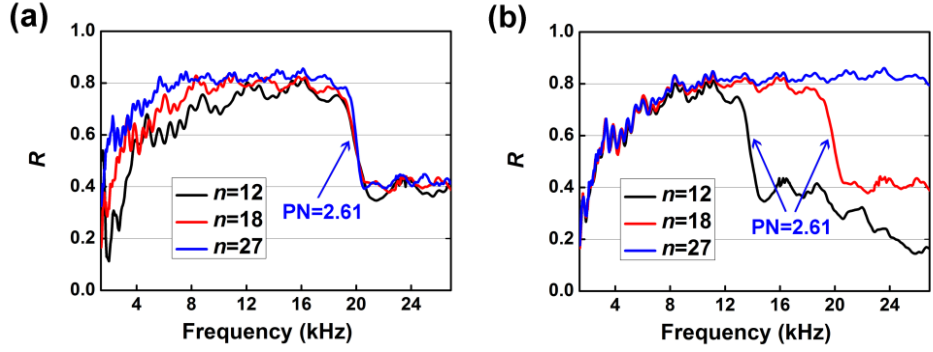

**FIG. 4** The reflection efficiency ($R$) spectra with division numbers ($n$) of 12, 18, and 27: (a) The period $d$ is constant; (b) The total width of ERUBAM is constant.

Furthermore, the influence of structural parameters on the efficiency and bandwidth of the ERUBAM has been investigated. The reflection efficiency of the ERUBAM is defined as $R = W/W_0$, where $W_0$ accounts for the average sound power incident on the ERUBAM, and $W$ is the average sound power of reflected wave across the specified cross-section of the 'Reflect Area' marked in Fig. 1(c). One has $W_0 = Sp_a^2/2\rho_0 c_0^2$ for plane wave, and $W = 1/2\int_s \text{Re}(p^*\boldsymbol{v})\cdot\boldsymbol{n}ds$ for reflected wave. Thus, $R$ represents the ratio between the average sound power across the reflected and incident cross-section, which can be used to quantitatively evaluate the performance of the ERUBAM. Figure 4 shows the reflection efficiency spectra simulated for three particular cases: $n=$ 12, 18, and 27, respectively. The results shown in Figs. 4(a) and 4(b) are calculated when the period $d$ and the width of ERUBAM are kept to be constant, respectively. For the ERUBAMs model with $n=18$, the effective bandwidths, which is defined as $R > 0.6$, is between 3 to 19.5 kHz (larger than 2 octaves). With a constant period $d$, as shown in Fig. 4(a), the low cutoff frequency ($f_{cl}$) of the ERUBAM is lowered with increasing $n$, while the high cutoff frequency ($f_{ch}$) remains invariant. On the other hand, with a constant ERUBAM width, $f_{ch}$ shifts to higher frequency, while $f_{cl}$ keeps constant when $n$ increases, as shown in Fig. 4(b).

The above results suggest that $f_{cl}$ is determined by the total widths of ERUBAM, whereas $f_{ch}$ is dominated by the period $d$. Since $\text{PN}=n/\beta=c_0/(2dgf)$, $f_{ch}$ is also determined by PN. Note that for different $n$, $f_{ch}$ is always smaller than the cutoff frequency of the waveguide $f_c=c_0/2d_0$ (e.g., for $n$=18, $f_{ch}=19.5\text{kHz}$ and $f_c=22.9\text{kHz}$). Thus, the existence of $f_{ch}$ lies in the fact that the phase separation is so rough to generate undesired reflected wave，and the critical value is PN=2.61. Consider an extreme condition of PN=2, that is, the phase alternately switches between 0 to $\pi$. In this case, the phase changes are symmetric from $x$ to $-x$ and $-x$ to $x$ are, leading to two symmetry reflected directions with corresponding angles of $\pm 45^\circ$, as shown in Fig. 5. Thus, in Fig. 4(a), the value of $R$ at PN=2 is nearly half of the value of $R$ at PN>2.61 because only half of the scattering beam reflects to the desired region.

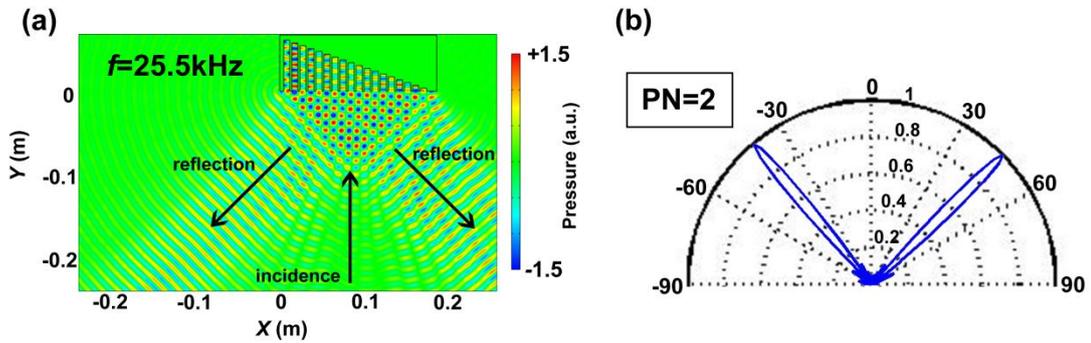

**Fig.5** (a) Scattering acoustic pressure field by FEM simulation at 25.5kHz (PN=2); (b) The $D(\theta)$ calculated by Eq. (9) at PN=2.

It is worth stressing that the reflection angle can be freely controlled by adjusting the gradient of groove depths, and more remarkably, versatile broadband manipulation can be realized such as sound focusing at an arbitrary position and non-diffractive beam with arbitrary convex trajectory, if *g* is chosen as a more complicated function of *x*. (See Supplementary Information).

In summary, we have proposed the concept of UBAM capable of yielding arbitrary manipulation on reflected wave within an ultra-broad bandwidth. A distinct property of yielding the extraordinary reflection by two-dimensional (2-D) UBAM structures has been designed theoretically and realized experimentally. Both the theoretical predictions and the experimental results reveal that a non-periodic phase array with the grating pattern could be used to achieve efficient control over the reflected sound wave without frequency limitation. An analytical model is developed to interpret this extraordinary phenomenon and provide a deeper insight to the underlying physics mechanism. By applying such an approach, it is also possible to achieve more complicated reflected phase profiles, such as non-diffractive beam with arbitrary convex trajectory, and vortex. The design can also be extended to three-dimensional case. The present work provides a very simple but high-efficient solution to realize UBAM by controlling the sound propagation distance in the grating pattern. Besides the apparent fundamental interests in the physics community, the UBAM structure may also have important applications a variety of practical situations where a special harness of acoustic wave is required such as medical application of ultrasound or field caustics engineering.

This work was supported by the National Basic Research Program of China (973 Program) (Grant Nos. 2010CB327803 and 2012CB921504), National Natural Science Foundation of China (Grant Nos. 11274168, 11174138, 11174139), and A Project Funded by the Priority Academic Program Development of Jiangsu Higher Education Institutions.

[†]These authors contributed equally to this work.
[*]Corresponding author.

jccheng@nju.edu.cn and liangbin@nju.edu.cn

Supplementary Information on "Ultra-Broadband Acoustic Metasurface for Manipulating the Reflected Waves"

Yi-Fan Zhu, Xin-Ye Zou, Rui-Qi Li, Xue Jiang, Juan Tu, Bin Liang, and Jian-Chun Cheng

**1. Extraordinary reflection at $30^o$ and $60^o$**

Figure S1 shows the $30^o$ and $60^o$ reflections corresponding to $g=0.25$ and $g=0.433$, respectively. The scattering acoustic fields are mapped at the frequencies of 7.071kHz, 12.162kHz, and 16.970kHz. The simulated results show that the ERUBAMs of $30^o$ and $60^o$ also have good broadband performance.

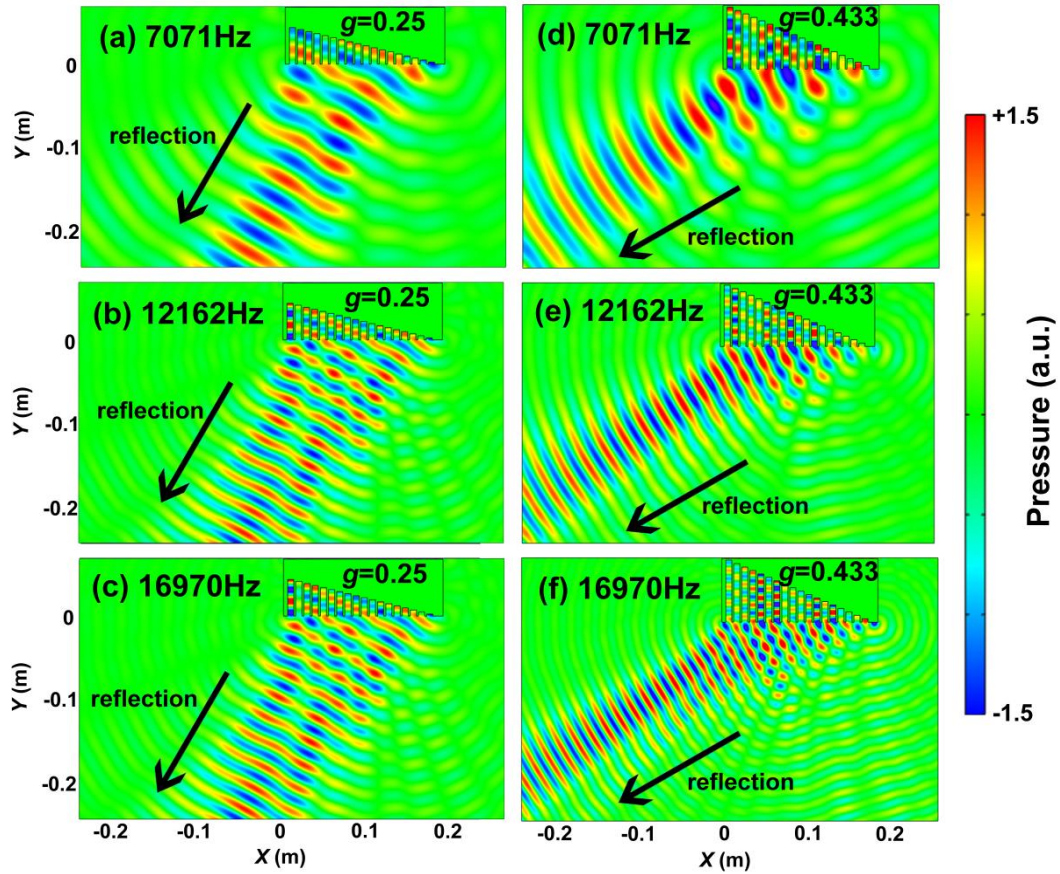

**FIG. S1** Scattering acoustic pressure fields of ERUBAM with $g=0.25$ in simulations: (a) 7.071kHz, (b) 12.162kHz and (c) 16.970kHz. Scattering acoustic pressure field of

ERUBAM with *g*=0.433 in simulations: (d) 7.071kHz, (e) 12.162kHz and (f) 16.970kHz.

## 2. Sound focusing

In order to focus at an arbitrary position $(x_0, y_0)$, a hyperboloidal phase profile is required

$$\phi_x = \frac{2\pi}{\lambda}\left[\sqrt{(x-x_0)^2 + y_0^2} - \sqrt{x_0^2 + y_0^2}\right]. \tag{S1}$$

Since $\phi_x = -2\pi \times (2h_x/\lambda)$, the depth of groove at different *x* locations is determined by

$$h_x = h_0 - \frac{1}{2}\sqrt{(x-x_0)^2 + y_0^2} \tag{S2}$$

where $h_0$ is a constant to insure that $h_x$ is a positive value. Figure S2(a)-(c) show the scattering fields of sound focusing model with the focal location (10cm, -19cm) at 7.071kHz, 12.162kHz and 16.970kHz, respectively. The focusing phenomenon is obvious.

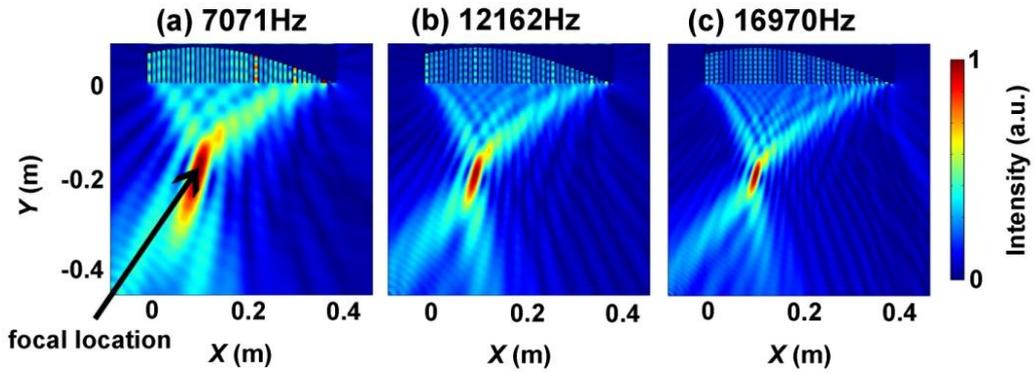

**FIG. S2** Simulated scattering acoustic pressure field of sound focusing at (a)7.071kHz, (b)12.162kHz, and(c) 16.970kHz.

## 3. Airy beam

Airy beam is another important beam. The Airy function has the asymptotic form expressed as

$$\mathrm{Ai}\left(-\frac{x}{x_0}\right) \approx \frac{1}{\sqrt{\pi}}\left(\frac{x}{x_0}\right)^{1/4} \sin\left[\frac{2}{3}\left(\frac{x}{x_0}\right)^{3/2} + \frac{\pi}{4}\right] \quad \text{(S3)}$$

where $x_0$ is the scalar factor. Thus, the depth satisfies

$$h_x = \frac{x_0}{2\sqrt{\pi}}\left(\frac{x}{x_0}\right)^{1/4} \sin\left[\frac{2}{3}\left(\frac{x}{x_0}\right)^{3/2} + \frac{\pi}{4}\right]. \quad \text{(S4)}$$

Figure S3(a)-(c) show the Airy beams with $x_0 = 0.2\,\mathrm{cm}$ at 7.071kHz, 12.162kHz and 16.970kHz, respectively.

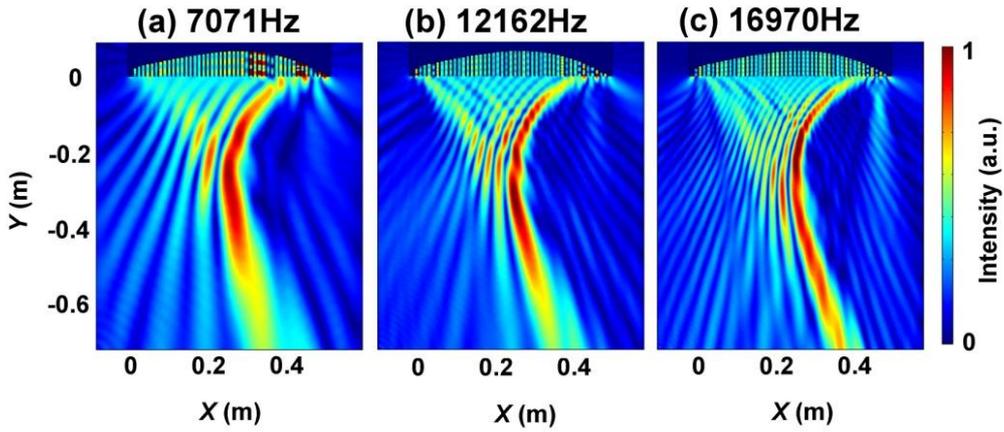

**FIG. S3** Simulated scattering acoustic pressure field of Airy beam model at (a)7.071kHz, (b)12.162kHz, and(c) 16.970kHz.

## 4. Non-diffractive beam with arbitrary convex trajectory

We have also realized UBAMs with generation of acoustic beam that propagates along arbitrary convex trajectory. As an example, a circular trajectory $(x+r)^2 + y^2 = r^2$ is designed. The phase should satisfy

$$\phi_x = \frac{2\pi}{\lambda}\left[\sqrt{(x+r)^2 - r^2} + r\arcsin\left(\frac{r}{x+r}\right)\right]. \tag{S5}$$

Thus, the depth $h_x$ is

$$h_x = h_0 - \frac{1}{2}\left[\sqrt{(x+r)^2 - r^2} + r\arcsin\left(\frac{r}{x+r}\right)\right]. \tag{S6}$$

Fig. S4(a)-(c) show the scattering fields at 7.071kHz, 12.162kHz and 16.970kHz, respectively.

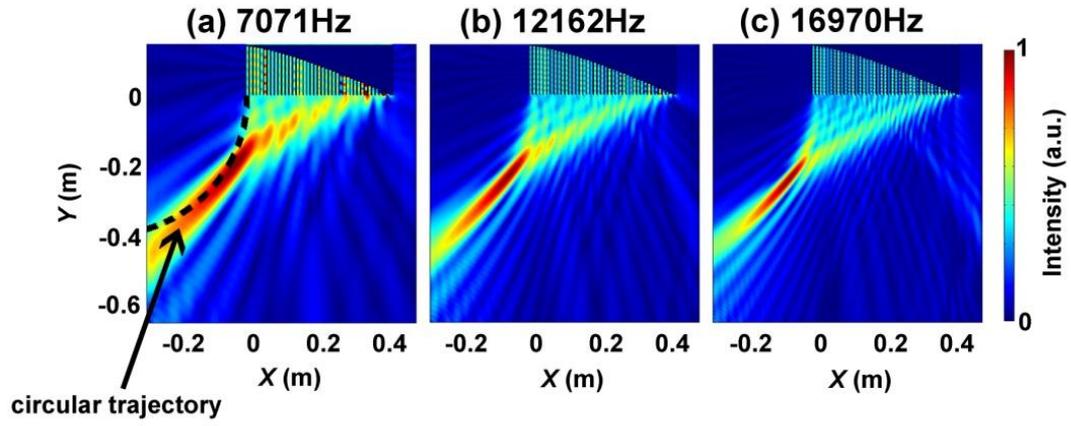

**FIG. S4** Simulated scattering acoustic pressure field of non-diffractive beam with arbitrary convex trajectory at (a)7.071kHz, (b)12.162kHz, and(c) 16.970kHz.